\documentstyle[12pt]{article}
\begin {document}
\begin{flushright} {	OITS 586}\\
{October 1995}
\end{flushright}
\vspace*{1cm}

\begin{center} {\large {\bf CHAOTIC BEHAVIOR OF PARTICLE PRODUCTION \\
\vspace*{.2cm}
IN BRANCHING PROCESSES}}
\vskip .75cm
 {\bf  Zhen CAO and Rudolph C. HWA}
\vskip.5cm
 {Institute of Theoretical Science and Department of Physics\\  University of
Oregon, Eugene, OR 97403}
\end{center}

\vskip .5cm
\begin{abstract}
The notion of chaotic behavior is examined for particle production in branching
processes.  Two types of branching are considered: non-Abelian gauge
interaction and an Abelian cascade model.  Properties of the production
processes are investigated by Monte Carlo stimulation.  The ``temporal''
behavior is studied by following the fluctuations in the multiplicities of each
generation as the branching evolves.  The ``spatial'' behavior is described in
terms of the fluctuations of the normalized factorial moments from event to
event.  The information dimension and a new entropy index are determined.
When all the measures are taken together, they collectively give a strong
suggestion that the QCD branching process is chaotic, while the Abelian cascade
model is not.
\end{abstract}
\vspace*{.8cm}
\section{Introduction}

Recently we have reported on the results of an investigation in the possible
signatures of chaos in branching processes \cite{zca}.  The aim is to determine
whether the nonlinear, non-Abelian dynamics of the quantum Yang-Mills field
possesses chaotic behavior.  Since in such a dynamical system the number of
degrees of freedom increases with time evolution, when the notion of time is
not
even well defined in the branching process, new measures of trajectory,
distance, entropy, etc.\ must be introduced.  We have found that the
perturbative QCD branching shows signs indicative of chaos, whereas a model
lacking the characteristics of QCD does not.  In this paper we give the details
of
our study.

In the case of classical non-Abelian dynamics special simplifying conditions
that reduce the equations of motion to manageable size have been considered
and chaotic solutions have been found \cite{sgm,sgm2}.  A more complete
investigation of the gauge equations has to be done on the lattice, and it has
been shown that the classical non-Abelian gauge theory generally exhibits
deterministic chaos, whose Lyapunov exponents can be numerically computed
[4-6].

The extension of the above classical problem to the quantum theory of
Yang-Mills field is extremely difficult.  The current state of knowledge about
quantum chaos hovers round semi-classical problems in which classical
trajectories are generalized to waves \cite{mt2,crgc}.  That is totally
inadequate
for treating quantum fields, the proliferation of which in a collision process
involves issues that are untouched in solving wave equations.  The first step
toward formulating a feasible program to attack the problem is to dissociate
the complications of nonperturbative QCD from the quantum dynamics of
non-Abelian fields.  In the perturbative domain where the QCD coupling is
small, the nature of nonlinear, non-Abelian dynamics is fully present so that
the signature of chaotic behavior in gauge theory should nevertheless show up
in hard processes.  That is an important step of simplification that renders
the
problem manageable.  More specifically, one can narrow the scope and focus
on branching processes with QCD splitting functions.  The issue then becomes
the search for measures that can reveal chaoticity in branching processes.

In such a search it is necessary to keep in mind the special features of QCD.
There are many processes that involve branching in real time, such as cell
division in biological systems.  The emission of photons by an accelerated
charge can be regarded as a branching process with sequential ordering.
However, for gluons reproducing gluons in the pure gauge theory, the time
variable plays no role in the description of a state with $n$ gluons whose
momenta are precisely specified.  In multiparticle production one works with
momenta (or their variants such as rapidity), since they are what can be
measured in a collision process.  Then in the momentum space the concept of
trajectory for a system of increasing number of gluons becomes ill defined.
Without trajectories the notion of distance between trajectories is untenable,
and the conventional method of examining chaos in classical nonlinear
problems is inapplicable.  It will therefore be necessary for us to look for
other
quantities that can describe the difference between states, which evolve from
what correspond to nearby initial conditions in the classical problem.

The major difficulty with the study of chaotic behavior of nonlinear gauge
dynamics is that the gauge fields are not directly measurable and that the
branching processes cannot be tracked experimentally.  In collisions at high
energies, where the question of chaos in gauge dynamics arises, only the
particles in the final states can be measured.  Thus to verify any theoretical
predictions about chaotic behavior, the loss of information at the end of the
branching process must be quantified and presented in a form suitable for
experimental determination.  To that end we shall introduce an entropy index
$\mu _q$, which is measurable, and describes the degree of fluctuation of the
final particles from event to event.

Before going into details, it may be helpful to state the general idea
underlying
the work.  In classical dynamics if the coordinates $q_\alpha (t)$ and momenta
$p_\alpha(t)$ specify the system, then the trajectory in phase space is well
defined in the familiar way.  In classical field theory the fields,
$E_\alpha(x,t)$
and $B_\alpha(x,t)$ say, form a field configuration defined over all space $x$
at any given time, and the change of the configuration as time evolves
specifies
a trajectory in a generalized sense.  For quantized fields one works in the
Fock
space so that the number of quanta becomes a variable that specifies an
important aspect of the state of the system.  The distance between two
trajectories in that case must therefore involve, among other quantities, the
difference between the numbers of quanta in the system.  Thus in particle
production at high energy where the multiplicity $n$ of produced particles is
an essential observable, the fluctuation of $n$ from event to event must be
viewed as the consequence of swings of trajectories that have almost the same
initial condition.  With that connection in mind it is natural to generalize
the
conventional treatment of nonlinear dynamics to an approach that places
emphasis on tracking the multiplicity fluctuations from event to event in high
energy collisions.  After describing the QCD dynamics of branching and a
simple $\chi$ model in Sec. 2, we treat the temporal behavior of the branching
processes in Sec. 3.  Then in Secs. 4 and 5 we consider the spatial properties
of the final state.  Concluding remarks are given at the end.

\section{Gluon Branching and the $\chi$ Model}

To carry out this investigation it is necessary to use computer simulation to
generate events of particle production through branching.  Only then is it
possible to study the nature of fluctuations of the final-state particles.  For
the
simulation we shall use algorithms based on two opposite types of dynamics.
One is, of course, the QCD dynamics.  For simplicity we shall focus on only the
pure gauge theory without quarks.  At the end of the evolution we shall
identify the partons as particles to avoid the complication of hadronization,
which is inessential to the question of whether the gluon branching dynamics
is chaotic.  The other is a cascade model, to be called the $\chi$ model, which
has none of the features of the QCD; in particular, it does not have infrared
and
collinear divergences.  It is studied in order to provide a contrast to the
gauge
theory so that our measures for chaoticity can be tested on these two
contrasting branching dynamics.

In both branching processes the initial parton has virtuality $Q^2$, and
successive branchings continue until the virtualities of all partons are $\leq
Q^2_0$.  In pure-gauge QCD the splitting function at each vertex of branching
is (for $g \rightarrow gg$)
  	\begin{eqnarray}  P(z)  = 6\left[ {1-z \over z}\, + \,{z
\over 1-z}\, +\,  z\left( 1 - z\right)\right] \quad ,
\label{1}
       \end{eqnarray}
where $z$ is the momentum fraction of the daughter parton in the frame
where the mother parton's momentum is 1.  In the $\chi$ model we keep only
the last term of (\ref{1})
   \begin{eqnarray}
P(z)  = 6 z \left( 1 - z\right)
  \label{2}
     \end{eqnarray}
so that it has no divergences at $z=0$ and $1$.  The presence of those
divergences is, of course, the source of complication for QCD that must be
treated carefully.

We follow Odorico's procedure \cite{odo} to develop the algorithm for Monte
Carlo simulation of parton shower in QCD.  Because of the soft gluon and
collinear divergences many partons are emitted at small $z$ and small angles
but are not resolvable.  The probability that a parton at $q^2 = Q^2$ can
survive
without emitting a resolvable parton until $q^2 = Q^2_0$ is the Sudakov form
factor \cite{odo,brw}
	   \begin{eqnarray}
\Delta  (Q,Q^2_0) = \mbox{exp}\left[ - \int^{Q^2}_{Q^2_0}  {dt^\prime \over
t^\prime} \, {\alpha_s (t^\prime) \over
2\pi} \int^{1-z_0(t^\prime)}_{z_0(t^\prime)}  dz \, P(z) \right]
\quad , \label{3}
     \end{eqnarray}
where the limits of the $z$ integration define what is meant by resolvable.
The first step of the simulation is to set $z_0 = Q^2_0/Q^2$ and calculate from
$\Delta(Q^2,Q^2_0)$ whether a parton starting from $Q^2$ evolves to $Q^2_0$
without branching.  If not, then the value of $q^2$ at which the branching
occurs is determined by solving
	 \begin{eqnarray}
{1 - \Delta(Q^2,Q^2_0)/\Delta(q^2,Q^2_0) \over 1 - \Delta(Q^2,Q^2_0)} = R_\#
\quad ,
\label{4}
     \end{eqnarray}
where $R_\#$ is a random number between 0 and 1.  With that $q^2$, the
value of $z$ is then generated in accordance to the distribution $P(z)$ given
in (\ref{1})	keeping only what is in the range $Q^2_0/q^2 \leq z \leq
1-Q^2_0/q^2$.  With that $z$, the daughter partons are assigned the maximum
virtualities $t_1 = zq^2$ and $t_2 = (1-z)q^2$, from which further evolutions
are carried out by repeating the above procedure starting with $t_1$ and
$t_2$.  When the virtuality of a descendant reaches $q^2 \leq Q^2_0$, then
that branch of the tree terminates.  We shall put $Q_0 = 1$ GeV, to be
definite.
In the simulation we shall set $N_c = 3$, $N_f = 0$ so that
   \begin{eqnarray}
\alpha _s (q^2) = 4\pi /11 \, \, \mbox{log} (q^2/\Lambda^2) \quad , \label{5}
     \end{eqnarray}
where $\Lambda^2$ will be set at 250 MeV.

We have also tried to simulate the parton shower using the algorithm of Weber
\cite{brw} and found that the result differs very little from that based on the
above algorithm of Odorico \cite{odo} for the type of measures we calculate.
Since Odorico's method is far more efficient, we have chosen to use it
throughout this work.

In the $\chi$ model, because of no infrared and collinear divergences, there is
no automatic $q^2$ evolution.  However, in order to compare its properties
with those of QCD, we introduce by hand $Q^2$ dependence of the branching
process by requiring  that at each vertex the two
daughter partons have virtualities $zq^2$ and $(1-z)q^2$, when the mother
virtuality is $q^2$, and $z$ is generated by using (\ref{2}) for $0 \leq z \leq
1$.  As with QCD, we require branching to continue successively until the
virtualities of all partons become $\leq Q^2_0$.  Since $q^2$ is not degraded
along a parton line, there are far more particles produced in the $\chi$ model
than in QCD for the same $Q^2$, but that is immaterial, since our measure of
chaoticity will not depend on $Q/Q_0$.  Clearly, the dynamics of branching in
the $\chi$ model is very different from QCD.  We use it to exemplify the
Abelian dynamics that has no infrared and collinear divergences.

\section{Temporal Behaviors of Branching}

Each branching process can be represented by a tree diagram, since
recombination of partons is not considered.  The vertices of the tree could be
ordered vertically in accordance to the values of $q^2$ of the mother partons.
Then there are many diagrams with the same topology that can describe
different evolution processes leading to the same number of particles at the
end, but belonging to different final states.  If we defer the considerations
of the
momenta of the final particles until the next section, we can simplify the
problem by focusing on only the  topology of the tree diagram.  In that way we
study the fluctuation of the particle multiplicity at the expense of ignoring
their
momenta.  What is gained is the possibility of defining a trajectory in the
multiplicity space for a branching process.

In emphasizing the topology of a diagram let us draw all diagrams in such a
way that all partons of the same generation are placed at the same level
regardless of their $q^2$.  An example is shown in Fig.\ 1.  Where the partons
are horizontally has no significance in that diagram.  Vertically, the
branching
points of the same generation are placed at the same level.  All partons that
reach the final state (represented by the dashed line)  have $q^2 \leq Q^2_0$.
No information concerning $q^2 > Q^2_0$ is carried by the vertical position of
the lines in the diagram.   Let $i$ denote the generation, staring with $i = 0$
for the initial parton at $q^2 = Q^2$.  Let $b_i$ denote the number of
branching
points at the ${\it i}$th generation.  In Fig.\ 1 we have $b_i = 1,2,3,3,1$ for
$i =
0,
\cdots , 4$.  We can define a vector ${\bf b} = (b_0, b_1, \cdots)$ with as
many components as there are nonvanishing $b_i$.  Thus for the diagram
in Fig.\ 1 we have ${\bf b} = (1, 2, 3, 3, 1)$.  If this description were to be
applied to an electron radiating photons, then a bremsstrahlung diagram would
have
${\bf b} = (1,1,1,\cdots)$.  On the other hand, for cell reproduction where
every
cell subdivides into two, then
${\bf b} = (1, 2, 4, 8, \cdots)$.  The description can be further simplified,
if we
define
     \begin{eqnarray}
x _i = \mbox{log}_2 \, b_i \label{6}
     \end{eqnarray}
and the corresponding vector ${\bf x} = (x _0, x _1, \cdots)$.  Since the
minimum $x_i$ is 0 and the maximum $x_i $ is $i$, the two extreme vectors
are ${\bf x} = (0, 0, 0, \cdots)$ which is like bremsstrahlung, and ${\bf x} =
(0,1,
2, 3, \cdots)$ which is like cell-reproduction.  They are shown in Fig.\ 2 by
the
thick lines, where $x_i$ is plotted against $i$.  All possible tree diagrams of
branching processes are represented by a line in between the two boundary
(thick) lines, as illustrated by the thin line in Fig.\ 2.  Such a line
specifies a
trajectory.

For finite $Q^2$ the branching cannot go without end, so there is always a
maximum $i$, which we denote by $i_{max}$.  The value of $i_{max}$ varies
from event to event, even if $Q^2$ is fixed for all events.  Two paths with the
same $i_{max}$ may differ for $0 < i < i_{max}$ , and the total numbers of
partons produced would be different.  In general, for any two trajectories
${\bf x}$ and ${\bf x}^\prime$ one can define $d_i = \left| x_i -
x_i^\prime\right|$ at every $i$ where $x_i$ and $ x_i^\prime$ exist
simultaneously.  Obviously, $d_i$ can be regarded as the distance between the
two trajectories.  Since two trajectories in $x$ space can cross, $d_i$ may
vanish at nonzero $i$.  Since $i$ is an integer, $x_i$ has discrete values.
Thus
the trajectories cannot be dense in the continuum space of $i$ and $x_i$, but
for the discrete support of the trajectories we may regard the collection of
all possible trajectories as being dense in the sense that a trajectory can
pass
through any allowed value of $(i,x_i)$.

To relate the description even more closely to the classical treatment of
chaos,
we now address the question of sensitivity to initial condition.  In classical
nonlinear dynamics one can specify the point in phase space where a
trajectory begins and vary that point $n$ in as small a neighborhood
$N_\varepsilon$ as one chooses.  In non-Abelian gauge dynamics or in the
$\chi$ model, the initial condition is that the initial parton has virtuality
$q^2
= Q^2$.  Being quantum mechanical, it is not necessary to vary $q^2$ in a small
neighborhood of $Q^2$.  Quantum fluctuation is sufficient to guarantee that
the final state of the branching process will vary from event to event, even if
all events start out at precisely the same $Q^2$.  Thus the relationship to the
classical consideration of initial condition is as follows.  Consider $\cal N$
random points in the neighborhood $N_\varepsilon$ of the initial point in
phase space of a classical trajectory.  Any such point can be mapped to the
beginning of a particular branching process.  A set of $\cal N$ events in the
Monte Carlo simulation all starting out at the same $Q^2$ then correspond to
$\cal N$ trajectories beginning in the neighborhood $N_\varepsilon$ in the
classical case.  With this correspondence between the two problems it is then
sensible to suggest that instead of studying the distance between two
neighboring trajectories, one should consider all $\cal N$ events in the
Monte Carlo simulation  and examine the mean deviation from the average of
the parton multiplicities.  Indeed, the study of fluctuations of observable
quantities will be the main theme of our approach to analyzing chaotic
behavior of branching processes.

In Fig.\ 3 we show schematically several possible trajectories of the branching
processes, all started out at the same $Q^2$.  Not only can trajectories that
have
the same $i_{max}$ be different, those with different $i_{max}$ are even more
dissimilar.  For the purpose of studying ``temporal'' behavior, which
corresponds to the dependence on the generation $i$ in the branching process,
only trajectories with the same $i_{max}$ can be compared.  Thus it is
necessary to know the distribution $P\left(i_{max}\right)$ of $i_{max}$ for all
events.  Fig.\  4 shows the result of $10^5$ simulations for each of the two
type
of branching processes discussed in the previous section.  The initial
virtualities are such that $Q/Q_0 = 10^3$ for QCD and 15 for the $\chi$ model.
These values are chosen so that the average multiplicities are comparable, as
we shall show below.  Both $P\left(i_{max}\right)$ are Gaussian-like
distributions, with the width being larger for QCD.  This is the first
indication of
more fluctuation for QCD as compared to the $\chi$ model.  In choosing a narrow
band of $i_{max}$ for further investigation, we select the shaded regions in
Fig.\
4 that are situated at the maximum of $P\left(i_{max}\right)$.

As mentioned before, time is not a well-defined variable in the branching
process.  It is deemphasized when we focus on the topololgy of the tree
diagrams.  The generation label $i$ carries only a rough notion of time, since
it
is well known, as in family genealogy, that a son can be younger than a
grandson.  However, there is one quantity that increases monotonically with
$i$; it is the parton multiplicity $n_i$
\begin{eqnarray}
n_i = 1 + \sum^{i-1}_{j=0} \, b_j = 1 + \sum^{i-1}_{j=0} \, 2^{x_j}  \quad ,
\label{7}
     \end{eqnarray}
even though $x_i$ may rise and fall with $i$. Define the average parton
multiplicity $\left< n_i \right> $ at generation $i$ by
\begin{eqnarray}
\left< n_i \right> =  {1 \over {\cal N}}\, \sum^{\cal N}_{e=1} \, n^e_i
(i_{max})
\quad ,
\label{8}
     \end{eqnarray}
where $n^e_i$ is the value of $n_i$ at $i$ for the {\it e}th event and the sum
is
over all ${\cal N}$ events having the same $i_{max}$ chosen.  This quantity
$\left< n_i \right>$ may be taken to play the role of time in the branching
process, although no linear dependence is implied.  In Fig.\  5 we show the
simulated results on $\left< n_i \right> $ as functions of $i$ for various
values
of $Q/Q_0$.  In both types of branching processes $\left< n_i \right>$
increases
monotonically with $i$, the rate being fastest at midrange of $i$.  The values
of
$Q/Q_0$ are chosen such that $\left< n_i \right>$ are roughly in the same range
for the two types.

For the measure that describes the fluctuation of the trajectories, and
therefore of the multiplicities, we use the normalized variance
\begin{eqnarray}
V_i =  {\left< n^2_i\right> - \left< n_i\right>^2
\over \left< n_i\right>^2} \quad , \label{9}
     \end{eqnarray}
which clearly gives a measure of the average ``distance'' between trajectories.
It is always positive, as a distance function should be.  In Fig.\  6 is shown
the
results of simulation for the two cases.  The sustained increase in the lower
half
of $i$ range for QCD is a distinctive feature that is not shared by the $\chi$
model.

A more transparent way of presenting this result is to plot $V_i$ vs  $\left<
n_i
\right>$, as shown in Fig.\  7.  It corresponds to plotting the distance $d(t)$
against $t$ in the classical problem, except that here  $\left< n_i \right>$ is
not
exactly $t$, but some representation of it.  Now, the behavior for QCD appears
universal, i.e., independent of $Q/Q_0$ over a wide range.  In the log-log plot
the behavior is approximately linear.  The same is not true for the $\chi$
model:  $V_i$ increases initially, then drops precipitously, as  $\left< n_i
\right>$ is increased.  Furthermore, the maximum decreases with increasing
$Q/Q_0$.   Thus the uncertainty in the parton multiplicity increases with
$\left< n_i\right>$ in the former case, but saturates and then decreases in the
latter case.  If that uncertainty can be regarded as a measure of chaos, then
the
QCD dynamics is chaotic, while the $\chi$ model is not.  We can express the
dependence of $V_i$ on  $\left< n_i\right>$ in the QCD case in the form of a
power law
  \begin{eqnarray}
V_i  \, \propto \, \left<n_i\right>^{\kappa}  \quad . \label{10}
     \end{eqnarray}
Then from Fig.\ \ 7 we find $\kappa$ to be $\simeq 0.4$.  Writing (\ref{10}) in
the form of an exponential
\begin{eqnarray}
V_i  \, \propto \, \mbox{exp}\left(\kappa \, {\rm ln} \left<n_i\right>\right)
\quad ,
\label{11}
     \end{eqnarray}
we may compare it to the definition of
Lyapunov exponent $\lambda$ in $d(t) \sim e^{\lambda t}$.  If ln$\left< n_i
\right>$ can be interpreted to correspond to $t$, then $\kappa$ may be said to
play the role of the Lyapunov exponent.  However, nothing firm can be proven
about these correspondences, so the above remarks should only be taken as a
possible orientation in interpreting the implication of the result obtained.

{}From what has so far been done, it is not possible for us to find a criterion
on
the magnitude of $\kappa$ to signify robust chaoticity.  Furthermore, $\left<
n_i
\right>$ and $V_i$ are aspects of the parton state at generation $i$ before the
completion of the branching processes.  Thus they are not experimentally
measurable.  For particle production in high energy collisions, the temporal
behavior of branching is primarily a theoretical problem.  For experimental
verification of any theoretical prediction, it is necessary to focus on the
final
state, which is the subject of the next two sections.

\section{Factorial Moments, Entropy, and Information Dimension}

In the final state the complete information about the partons, now identified
as observable particles, is registered by their momenta.  If the branching
trajectories never change, then there is certainty in where to find those
particles in the momentum space, and the entropy (to be defined below) is zero,
corresponding to no loss of information.  However, quantum fluctuation alone is
sufficient to cause fluctuation in the final-state momenta, so entropy is not
expected to vanish.  In multifractal analysis of complex patterns, the
information dimension is a compact way of summarizing the dependence of the
entropy on resolution scale \cite{hgs}.  Since multifractal analysis of the
multiplicity fluctuation in high-energy collisions has been proposed
\cite{hw2,hw3} and carried out \cite{cbc,ddk2}, it is natural to examine
the information dimension for the branching processes considered here

If $p_j$ denotes the fraction of particles in an event that fall into the $j$th
bin
of size $\delta$, then in the limit of many such bins in the system under
study,
the entropy is defined by \cite{hw3}
	\begin{eqnarray}
 S = -\sum_j \, p_j \, {\rm ln} \, p_j \quad , \label{12}
     \end{eqnarray}
and the information dimension $D_1$ is \cite{jf}
      \begin{eqnarray}
D_1 = -\lim_{\delta \rightarrow 0} S/{\rm ln} \, \delta  \quad . \label{13}
     \end{eqnarray}
The study of self-similar behavior of particle production has been greatly
facilitated by the use of normalized factorial momenta $F_q$, suggested by
Bia\l as and Peschanski \cite{bp}, who first showed that the statistical
fluctuations are filtered out by those moments.  However, since $F_q$ is
defined
for integer $q \geq 2$, it cannot be used to determine the information
dimension, which involves the derivative of $F_q$ with repect to $q$ at $q=1$.
For that reason the $G_q$ moments for all real $q$, positive and negative, were
introduced for analyzing the multifractal structure of particle production data
\cite{hw2,hw3,ddk2}.  Their drawback is that the statistical fluctuations must
be taken out of $G_q$ explicitly by hand.  More recently, a method for
continuing $F_q$ to noninteger $q$, while maintaining its virtue of not being
contaminated by statistical fluctuations, has been developed
\cite{rch}.  We shall therefore make use of it to determine $D_1$.

Before describing the details of the calculations, let us first make clear the
space in which the self-similar behavior is examined.  With $z_i$ denoting the
momentum fraction of an {\it i}th generation daughter parton at a branching
vertex, where the mother parton has momentum 1, the momentum fraction of a
final particle is then
	\begin{eqnarray}
x = \prod_i \, z_i \quad . \label{14}
     \end{eqnarray}
where the product is taken over all generations of a particular path in the
branching tree, leading from the initial parton to the final particle under
consideration.  Since all $z_i$ are known in a simulated event, the values of
$x$
for all particles can be calculated; they are all crowded in the $x=0$ region,
since
their sum is 1.  This is especially true for QCD branching because of the soft
gluon divergence in (\ref{1}).  If we expand the $x=0$ region by using the
variable
$\zeta = - \log _{10}x$, we get a broad Gaussian-like distribution, as shown is
Fig.\  8, with the peak at $\zeta = 3$ corresponding to $x = 10^{-3}$.  The
situation is not as bad for the $\chi$ model, but still most particles are
found in
the small $x$ region, as shown in Fig.\  9.  This highly uneven distribution of
$x$
is inappropriate for data analysis that involves the partition of the space
into
small bins.

One can define another variable $X$, in terms of which the distribution is
much smoother \cite{bg}.  Let the inclusive distribution in $x$, averaged over
many events, be $\rho (x)$.  Then the cumulative variable $X$ is defined by
		\begin{eqnarray}
X(x) = \int_{x_1}^x \rho(x')\,dx' / \int_{x_1}^{x_2} \rho(x')\,dx' \quad,
\label{15}
     \end{eqnarray}
where $x_1$ and $x_2$ are two extreme points in the distribution $\rho (x)$,
between which $X$ varies from 0 to 1.  In terms of $X$ the inclusive
distribution $\rho (X)$ is constant.  For QCD branching we shall take $\rho
(x^\prime)$ to be the $\zeta$ distribution as in Fig.\  8; for the $\chi$ model
$\rho
(x^\prime)$ will be just the $x$ distribution as in Fig.\  9.  An illustration
of the
$\rho (X)$ distribution for QCD is shown is Fig.\  10 after simulating
$5\times10^4$ events at $Q/Q_0 = 10^3$.  Note the expanded scale along the
vertical axis.

Now, in $X$ space we divide the interval $0 \leq X \leq 1$ into $M$ bins of
width $\delta = 1/M$.  Let $n_j$ be the number of final particles in the $j$th
bin in any given event.  The factorial moment of $q$th order is
    \begin{eqnarray}
f_q(M) = M^{-1} \sum^M_{j = 1} n_j (n_j - 1) \cdots (n_j -
q + 1) \quad . \label{16}
     \end{eqnarray}
The normalized factorial moment after averaging over all events is
   \begin{eqnarray}
F_q = \left< f_q\right> / \left< f_1\right>^q \quad .
 \label{17}
     \end{eqnarray}
Since $\left< f_q\right>$ is the average of $n_j!/(n_j - q)!$ over all bins and
over all events, we may write it as
   \begin{eqnarray}
\left<f_q\right> = \sum^{\infty}_{n=q} \, {n! \over (n - q) !} \, P_n \quad .
 \label{18}
     \end{eqnarray}
where $P_n$ is the multiplicity distribution in a bin.  In general, $P_n$ may
be
expressed as a convolution of the statistical $(S)$ and the dynamical $(D)$
contributions to the multiplicity fluctuations:
   \begin{eqnarray}
P_n =  S \otimes D  = \int^{\infty}_0 dt \, {t^n \over n!}\,\, e^{-t}\, D(t)
\quad ,
 \label{19}
     \end{eqnarray}
where a Poissonian distribution has been used for $S$.  Putting (\ref{19}) into
(\ref{18}) yields
   \begin{eqnarray}
\left<f_q\right> = \int^{\infty}_0 dt \, t^q \,  D(t) \quad . \label{20}
     \end{eqnarray}
This is the standard moment of the dynamical distribution that has no
statistical
contamination \cite{bp}.  If $D(t) = \delta (t - \bar n)$, so that $P_n$ is
Poissonian, then $\left<f_q\right> = \bar n ^q$ and $F_q = 1$.  Any deviation
of
$F_q$ from 1 is an indication of the presence of dynamical fluctuation.

Phenomenologically, one does not have access to $D(t)$.  Using $P_n$ as input
from experiment or from simulation, one can determine
$\left<f_q\right>$, but only for integer $q$.  One may replace the factorials
in
(\ref{18}) by gamma functions in order to continue (\ref{18}) to noninteger
$q$ \cite{blaz}.  But that procedure does not result in $F_q = 1$ for Poisson
distribution for all $q$.  In Ref. \cite{rch} a continuation procedure is
developed that guarantees $F_q = 1$ for all $q$ when $D(t)$ is trivial.  That
method will not be described here, but will be used to determine $F_q$ for our
branching processes.

For notational brevity let us write the RHS of (\ref{20}) as $\left<
t^q\right>_D$.  Then we have
\begin{eqnarray}
F_q = \left<\left(  t/ \left< t\right>_D \right)^q\right>_D =  \left<\left(  pM
\right)^q\right>_D
 \quad , \label{21}
     \end{eqnarray}
where $p$ is the fraction of event multiplicity in a bin, as in (\ref{12}).
Thus
from (\ref{12}) we get
  \begin{eqnarray}
S = -M\left<  p \, {\rm ln} \, p \right>_D = \left.{\rm ln} \, M - {d \over dq}
\,
{\rm ln}  \, F_q\right| _{q=1}\quad .
\label{22}
     \end{eqnarray}
If $F_q$ has a power-law behavior near $q = 1$, i.e.,
   \begin{eqnarray}
F_q \, \propto \, M^{\varphi_q} \quad , \label{23}
     \end{eqnarray}
it then follows from (\ref{13}) that in the large $M$ limit
   \begin{eqnarray}
D_1 = 1 - \left.{d\varphi_q \over dq}\right|_{q=1} \quad . \label{24}
     \end{eqnarray}
Note that this relationship is derived under the assumption that (\ref{22}) is
meaningful, which demands that $F_q$ is well defined in a range of $q$
around $q=1$ and that $dF_q/dq$ contains only dynamical information
without spurious contribution arising from improper continuation procedure
\cite{rch}.

If dynamically the bin multiplicity is evenly distributed among all $M$ bins so
that $D(t) = \delta(t - t_0)$, then $F_q = 1$ for all $q$ (including
non-integer
$q$) and (\ref{22}) gives $S = {\rm ln} M$.  On the other hand, if dynamically
only a fraction $a$ of the $M$ bins are populated, the rest empty, i.e.
   \begin{eqnarray}
D(t) = a \delta \left({t - t_1}\right) + (1 - a) \delta(t) \quad , \label{25}
     \end{eqnarray}
then $\left< f_q\right> = at^q_1$ and $F_q = a^{1-q}$ for all $q$.  Hence,
(\ref{22}) yields $S = {\rm ln}(aM)$.  If, in particular, only one bin is
non-empty, i.e., $a = 1/M$, then $S=0$.  Thus (\ref{25}) offers a simple way of
seeing how the entropy increases from $0$ to ${\rm ln}M$, as the fraction of
non-empty bins increases from $1/M$ to $1$.  If there is only one non-empty
bin, the dynamics is like one with a classical trajectory that has no
dispersion in
where the particles are to be found in the momentum space.  Without loss of
information the entropy is zero.  If there is maximum dispersion corresponding
to all bins being equally likely to be populated, then the entropy is maximum.

We now apply this method of analysis to our simulated branching processes.
First, we show  in Fig.\ 11 the multiplicity distributions in QCD for various
bin
sizes $(\delta = 1/M)$ after $10^5$ events.  It should be understood that Fig.\
10 shows the inclusive $X$ distribution after summing over all events.  But
event by event the fluctuation in $X$ space is much greater.  Upon dividing the
interval
$0 \leq X \leq 1$ into $M$ bins and counting the multiplicity in each bin, one
then obtains $P_n(M)$ after averaging over all the events.  The degree of
fluctuation around the decreasing $\left< n\right>$, relative to that
average, as
$M$ is increased, is what is measured by the normalized factorial moments
$F_q$ defined by (\ref{17}) and (\ref{18}).  This is unlike our theoretical
calculation based on (\ref{20}) and (\ref{25}) because $D(t)$ is not known
explicitly in the simulation.  The results are shown in Fig.\ 12 for both QCD
and
$ \chi $ model.  Since $F_{q=1} = 1$ by definition, we see that $F_q$ increases
with $q$ in QCD, but decreases with increasing $q$ in the $\chi$ model.  It
means that the distributions $P_n(M)$ for QCD are wider than Poisson, but for
the
$\chi$ model they are sub-Poissonian.  Fig.\  12 also shows that there is very
little $M$ dependence in either case.  For non-integer $q$ values a substantial
amount of work is needed to continue $F_q$ according to the method of
\cite{rch}.  The result is shown in Fig.\  13.  It is clear that for $q >1$ all
$F_q$
are greater in QCD than in the $\chi$ model.  It corresponds to more
fluctuations in QCD than in the
$\chi$ model, a result that is consistent with the finding of the previous
section.

It is also evident from Fig.\  13 that $\left. dF_q/dq\right|_{q=1}$ is
essentially independent of $M$ in both cases. Whereas $S$ is still dependent on
$M$ according to (\ref{22}), the $M$ independence of $\left. d {\rm ln}
F_q/dq\right|_{q=1}$ requires, on account of (\ref{23}), that $\left. d \varphi
_q/dq \right|_{q=1} = 0$.  This together with (\ref{24}) entails
   \begin{eqnarray}
D_1 = 1  \label{26}
     \end{eqnarray}
in both cases.  Thus there is no interesting multifractal  property.  This
particular approach to finding a useful signature of chaotic behavior ends in
failure.

The smoothing of the phase space by use of the $X$ variable in place of the
momentum fraction variable $x$ has undoubtedly contributed to eliminating
some aspect of the fractal structure.  But what has not been removed is the
more interesting consequences of the dynamics.  The triviality of $D_1$ states
only that there is no nontrivial self-similarity in the fluctuations measured.
The latter are quantified by $F_q$, which involves both the bin and event
averages of the factorial product in (\ref{16}).  The result suggests that some
of
the fluctuations have been lost by the averaging process.  In the next section
we develop a scheme to recapture that which has been lost.

\section{Entropy Index}

We know that event by event there are large fluctuations in where the
produced particles are in the $X$ space.  What we want to do now is to find a
way to register the fluctuation (from event to event) of the fluctuation (in
the
$X$ distribution).

Let $F^e_q$ denote the normalized factorial moment of the {\it e}th event:
   \begin{eqnarray}
F^e_q = f^e_q/\left(f^e_1\right)^q \quad , \label{27}
     \end{eqnarray}
where $f^e_q$ is as defined in (\ref{16}) for the {\it e}th event.  Clearly,
$F^e_q$ changes with $e$ for fixed $q$ and $M$.  In Fig.\  14 we show the
distribution $P(F_q)$ after $10^5$ simulated events of QCD branching, the label
$e$ being omitted.  For clarity, only $q=2$ and $3$ are exhibited for a wide
range of $M$ values.  At small $M$ the distribution has a peak at
$F_q \approx 1$.  Indeed, if $M=1$ and event multiplicity is large, then $F_q$
should be very nearly 1 for all events.  At large $M$ a peak at $F_q = 0$ is
developed, especially at high $q$, because the bin multiplicity for small
$\delta$ is small and may in many events be less than $q$ for all bins, in
which
cases $F^e_q = 0$ .  This large fluctuation in $P(F_q)$ is what we want to
capture, and would be lost if $F_q$ is averaged over all events.  Clearly we
should take the moments of $F_q$ before performing the average.

We may think of $F^e_q$ as the horizontal moments in the $X$ space, and then
define the vertical moments in the event space as follows:
    \begin{eqnarray}
\left< F^p_q\right> = {1\over {\cal N}} \, \sum^{\cal N}_{e=1} \left(F^e_q
\right)^p \quad ,
\label{28}
     \end{eqnarray}
where ${\cal N}$ is the total number of events and $p$ is a positive real
number
not restricted to integers.  The normalized moments of moments are
		\begin{eqnarray}
C_{p, q}(M) = \left< F^p_q(M) \right>/\left< F_q(M)
\right>^p \quad . \label{29}
     \end{eqnarray}
If $C_{p, q}(M)$ has a power-law behavior in $M$, i.e.,
   \begin{eqnarray}
C_{p, q}(M) \propto M ^{\psi_q (p)} \quad , \label{30}
     \end{eqnarray}
then we can define
   \begin{eqnarray}
\mu_q = {d\over dp}\psi_q (p)|_{p=1} \quad . \label{31}
     \end{eqnarray}
which will be referred to as the entropy index.

To see the connection between $\mu_q$ and entropy, let us define
    \begin{eqnarray}
P_q^e = F_q^e / \sum^{\cal N}_{e=1} F_q^e  \label{32}
     \end{eqnarray}
and a (new) entropy in the event space
   \begin{eqnarray}
S_q = -\sum^{\cal N}_{e=1} P_q^e \,  {\rm ln} \, P_q^e \quad . \label{33}
     \end{eqnarray}
Furthermore, define the moments
    \begin{eqnarray}
H_{p,q} = \sum^{\cal N}_{e=1} (P_q^e)^p \quad , \label{34}
     \end{eqnarray}
which are related to $S_q$ by
    \begin{eqnarray}
S_q =  - {d\over dp}\, {\rm  ln}\, H_{p,q}|_{p=1} \quad . \label{35}
     \end{eqnarray}
On the other hand,  $H_{p,q}$ can be related to $C_{p,q}$ by
     \begin{eqnarray}
C_{p,q} = {\cal N}^{p-1} \, H_{p,q} \quad , \label{36}
     \end{eqnarray}
so we have
     \begin{eqnarray}
{d\over dp}\, {\rm ln}\,  C_{p,q}|_{p=1} = {\rm ln}\,  {\cal N} - S_q
\quad.
\label{37}
     \end{eqnarray}
In then follows from (\ref{30}) and (\ref{31}) that
     \begin{eqnarray}
S_q = {\rm ln} \,  ( {\cal N}  M^{-\mu_q} )\quad, \label{38}
     \end{eqnarray}
apart from a possible additive term that is independent of ${\cal N}$ and $M$.

This entropy, as defined in (\ref{33}), is clearly different from the entropy
$S$
defined in (\ref{12}).  To emphasize that $S_q$ is defined in the event space,
we
may call it eventropy.  We can think of the event space as a one dimensional
space with ${\cal N}$ sites.  At each site we can register a number $F_q^e$.
If
$F_q^e$ is the same at each site, then $P_q^e = 1/{\cal N}$, and $S_q =
{\rm ln} {\cal N}$. We should think of this as being highly disordered in the
event space, since $F_q^e$ is spread out uniformly over all space.  The larger
the
number of events, the larger is the eventropy.  This is very similar to the
situation where $S= {\rm ln} M$, when the bin multiplicity is uniformly
distributed in the $X$ space that has $M$ bins.  A branching dynamic that
results in the same $F_q^e$ for every event does not fluctuate in the branching
processes.  It corresponds to nearby trajectories staying nearby throughout.
In
short, the dynamics is not chaotic.  For $S_q$ to be ${\rm ln}{\cal N}$, we see
from (\ref{38}) that $\mu _q$ must vanish.  Thus small $\mu _q$ corresponds
to large eventropy, which in turn implies no chaotic behavior.

On the other hand, if we consider the other extreme where all $F_q^e$ are zero
except one event $e^\prime$, then $P_q^e = \delta _{ee^\prime}$ and $S_q = 0$.
This is highly ordered in the event space, but the fluctuation of $F_q^e$ from
zero to a nonzero value is large.  Generally speaking, if the distribution
$P(F_q)$
is broad, the fluctuation is large, initially nearby trajectories become widely
separated in the final states of different events, and the dynamics is chaotic.
 In
order for the eventropy to be small, $\mu _q$ must be large.  Thus large
entropy index implies chaotic behavior.

In Fig.\ 15 the results of our simulation are shown for $q=2$, $3$, $4$ and $p=
0.5$, $1.0$, $1.5$, and $2.0$.  From the log-log plots one can identify
approximate scaling behavior.  We used the region $M=5-20$ to determine the
value of
$\mu _p$, for which $C_{p,q}$ for an incremental region of $p$ around $p=1$
has to be examined.  We obtain
\begin{eqnarray*}
\mu _q^{(\rm QCD)} &= \quad 0.0061,\quad 0.054, \quad 0.23 \quad (q = 2, 3,
4)
\quad,
\nonumber\\
\mu _q^{(\chi)} &= \quad 0.0014, \quad 0.010, \quad 0.046 \quad (q = 2, 3, 4)
\quad.
\end{eqnarray*}
This result is plotted in Fig.\ 16.   Clearly, $\mu _q^{(QCD)}$ is much larger
than $\mu _q^{(\chi)}$.  It is suggestive that the QCD dynamics is chaotic,
while
the $\chi$ model is not.  However, at this point we have no quantitative
criterion on how large $\mu _q$ must be in order to be chaotic.  In classical
nonlinear problems there is also no criterion on how long the time lapse must
be in order for the distance between nearby trajectories to become sufficiently
far apart to qualify for divergent behavior.  Nevertheless, the positivity of
the
Lyapunov exponent is a simple condition, the counterpart of which for the
entropy index is lacking.

\section{Conclusion}

We have explored various ways of measuring chaoticity of branching
processes.  Because the problem is far more complicated than a classical
nonlinear problem, the search has not been a straightforward extension of the
conventional ideas.  Our approach has been based on a simple premise:  if a
branching process is chaotic, then it should be hard to predict what the likely
outcome of an event would be, given the knowledge of the result of a previous
event.  It means that there are large fluctuations from event to event.

To be more concrete, we have focused first on the temporal development of
branching.  The variance $V_i$ of the multiplicities at a particular generation
$i$ can be regarded as a measure of the mean distance between ``trajectories''.
We have found that $V_i$ increases with $\left< n_i \right>$ in a power-law
fashion for QCD branching, but not for the $\chi$ model, an example of Abelian
branching dynamics.  That is the first indication that the QCD dynamics has the
characteristics of being chaotic.  We have not investigated the behavior of the
higher moments, $V_i$ being related to only the second moment.  It is probably
safe to predict that the higher moments would increase even more rapidly with
$\left< n_i \right>$.

For fluctuations in the final state where experimental observations are
feasible,
we have considered $F_q$, $D_1$, $C_{p,q}$, and $\mu_q$ as possible measures
of chaoticity.  It turns out that the information dimension is not effective,
but
the entropy index can be.  What emerges is a realization that one should study
the phase space and the event space simultaneously.  More specifically,
consider a two-dimensional lattice, where the horizontal axis has $M$ sites,
corresponding to the $M$ bins in the $X$ space, and where the vertical axis has
${\cal N}$ sites, corresponding to the ${\cal N}$ events in the event space.
For
every event there is a horizontal array of $M$ numbers, indicating the particle
multiplicities in the $M$ bins.  The moments of moments $C_{p,q}$ summarize
the distribution on this whole lattice without significant loss of information.
 The
entropy index $\mu_q$ extracts from this lattice of numbers a simple numerical
quantity that can characterize the degree of fluctuations on the lattice.  The
larger $\mu_q$ is, the more chaotic the system is.  Unfortunately, we do not
have at this point a criterion on the threshold value of $\mu_q$, above which
chaotic behavior will definitely occur.  Further investigation on this aspect
of the
problem  is undoubtedly warranted.

Each of the features studied reveals some aspect of the chaotic behavior of QCD
dynamics.  Taken together collectively, they present a strong contrast from the
results of the $\chi$ model.  One may wonder what the mechanisms are in the
branching processes that can lead to such diverse outcomes.  Our view is that
in
QCD branching there is a tension between two opposing forces that is absent in
the $\chi$ model.  The collinear divergence implies  a strong preference
for small angle emission of  partons in the branching process.  But if the
emission angle is too small to be resolved, there is no branching.  Since
evolution without branching is suppressed by the Sudakov form factor,
there is a competition between evolution by emitting a resolvable gluon and a
preference for emitting a collinear unresolvable gluon.  It is this tension
that
leads to large fluctuations in the number of branching points and hence to
chaotic behavior.  The $\chi$ model has no collinear divergence and no
$q^2$ evolution, so there is no tension
to cause large fluctuations.  What we learn here may lay the foundation for
finding ways to treat nonperturbative processes, for which chaotic behavior
may play a crucial role in the evolution of a large system of quarks and
gluons,
such as those produced in a heavy-ion collision at high energy.

\begin{center}
\subsection*{Acknowledgment}
\end{center}

One of us (RCH) is grateful to C.\ B.\ Chiu, K.\ Geiger, B.\ L.\ Hao, S.\ G.\
Matinyan, and B.\ M\"uller for helpful discussions.  This work was supported in
part by the U.S. Department of Energy under Grant No. DE-FG06-91ER40637.

\newpage
\begin{center}
\section*{Figure Captions}
\end{center}

\begin{description}
\item[Fig. 1]A tree diagram for {\bf b} = (1, 2, 3, 3, 1)
\item[Fig. 2]A trajectory of $x_i$
\item[Fig. 3]Several possible trajectories.
\item[Fig. 4]Distribution of $i_{max}$ for QCD and the $\chi$ model.
\item[Fig. 5]Simulated results for the average multiplicities at various
generations
$i$ for fixed values of $Q/Q_0$ indicated by the numbers.
\item[Fig. 6]Simulated results for the normalized variances $V_i$ for the same
parameters as in Fig. 5.
\item[Fig. 7]Temporal behaviors depicted by $V_i$ vs $\left< n_i\right>$.
\item[Fig. 8]Momentum distribution in terms of $\zeta = \log x$ for QCD
branching.
\item[Fig. 9]Momentum distribution in terms of $x$ for the $\chi$ model.
\item[Fig. 10]Momentum distribution in terms of the cumulative variable $X$ for
QCD branching.
\item[Fig. 11]Multiplicity distributions in bins of size $\delta = 1/M$ for QCD
branching.
\item[Fig. 12]$F_q$ vs $M$ for $q = 2-5$.
\item[Fig. 13]$F_q$ vs continuous $q$.
\item[Fig. 14]Distributions of $F^e_q$ calculated event by event.
\item[Fig. 15]Moments of moments $C_{p,q}$ vs $M$ for $p = 0.5 - 2.0$ and $q =
2,3,4$.

\end{description}


\newpage

\begin{thebibliography}{00}

\bibitem{zca} Z.\ Cao and R.\ C.\ Hwa,  Phys.\ Rev.\ Lett.\  {\bf 75}, 1268
(1995).

\bibitem{sgm}S.G. Matinyan {\it et. al.}, Sov. Phys. JETP {\bf 53}, 421
(1981).

\bibitem{sgm2}S.G. Matinyan, Sov. J. Part. Nucl. {\bf 16}, 226 (1985) and
other references quoted therein.

\bibitem{mt}B. M\"{u}ller and A. Trayanov,  Phys. Rev. Lett. {\bf 68}, 3387
(1992).

\bibitem{bgmt}T.S. Bir\'{o}, C. Gong, B. M\"{u}ller and A. Trayanov, Int. J.
Mod. Phys. {\bf C5}, 113 (1994).

\bibitem{cg}C. Gong, Phys. Lett. B{\bf 298}, 257 (1993), Phys. Rev. D {\bf
49}, 2642 (1994).

\bibitem{mt2}M Tabor, {\it Chaos and Integrability of Nonlinear Dynamics}
(Wiley, N.Y., 1989).

\bibitem{crgc}{\it Quantum Chaos}, edited by H.A. Cerdeira {\it et al.}
(World Scientific, Singapore, 1991).

\bibitem{odo}E. Odorico, Nucl. Phys. B {\bf 172}, 157 (1989).

\bibitem{brw}B.R. Webber, Ann. Rev. Nucl. Part. Sci. {\bf 36}, 253 (1986).

\bibitem{hgs}H.G. Schuster, {\it Deterministic Chaos}, 3rd edition
(Physik-Verlag,
Weinheim, 1995).

\bibitem{hw2}R.\ C.\ Hwa, Phys.\ Rev.\ D{\bf 41}, 1456 (1990); C.\ B.\ Chiu and
R.\ C.\ Hwa, Phys.\ Rev.\ D{\bf 43}, 100 (1991).


\bibitem{hw3}For a review see R.\ C.\ Hwa,  in {\it Quark-Gluon Plasma}, edited
by
R.\ C.\ Hwa (World Scientific, Singapore, 1990).

\bibitem{cbc}
C.\ B.\ Chiu and R.\ C.\ Hwa, Phys.\ Rev.\ D{\bf 45}, 2276 (1992).

\bibitem{ddk2}For a review see E.\ A.\ De Wolf, I.\ M.\ Dremin and W.\ Kittel,
Phys.\ Rep.\ C (to be published).

\bibitem{jf}
J.\ Feder, {\it Fractals} (Plenum Press, N.\ Y.\ and London, 1988).

\bibitem{bp} A. Bia\mbox{\l}as and R. Peschanski, Nucl. Phys. B {\bf 273},
703 (1986); B {\bf 308}, 867 (1988).

\bibitem{rch}R.C. Hwa, Phys. Rev. D {\bf 51}, 3323 (1995).

\bibitem{bg}A. Bia\mbox{\l}as and M. Gardzicki, Phys. Lett. {\bf B252}, 483
(1990).

\bibitem{blaz} M.\ Blazek, Phys.\ Lett.\  B{\bf 247}, 576 (1990); P.\ Duclos
and
J.-L.\ Meunier, Zeit.\ Phys.\ C{\bf 64}, 295 (1994); I.\ M.\ Dremin, Pisma v.\
ZhETF{\bf 59} 561 (1994);  Uspekhi Fiz.\ Nauk {\bf 164}, 785 (1994); JETP
Lett.\ {\bf 59}, 585 (1994).

\end{thebibliography}
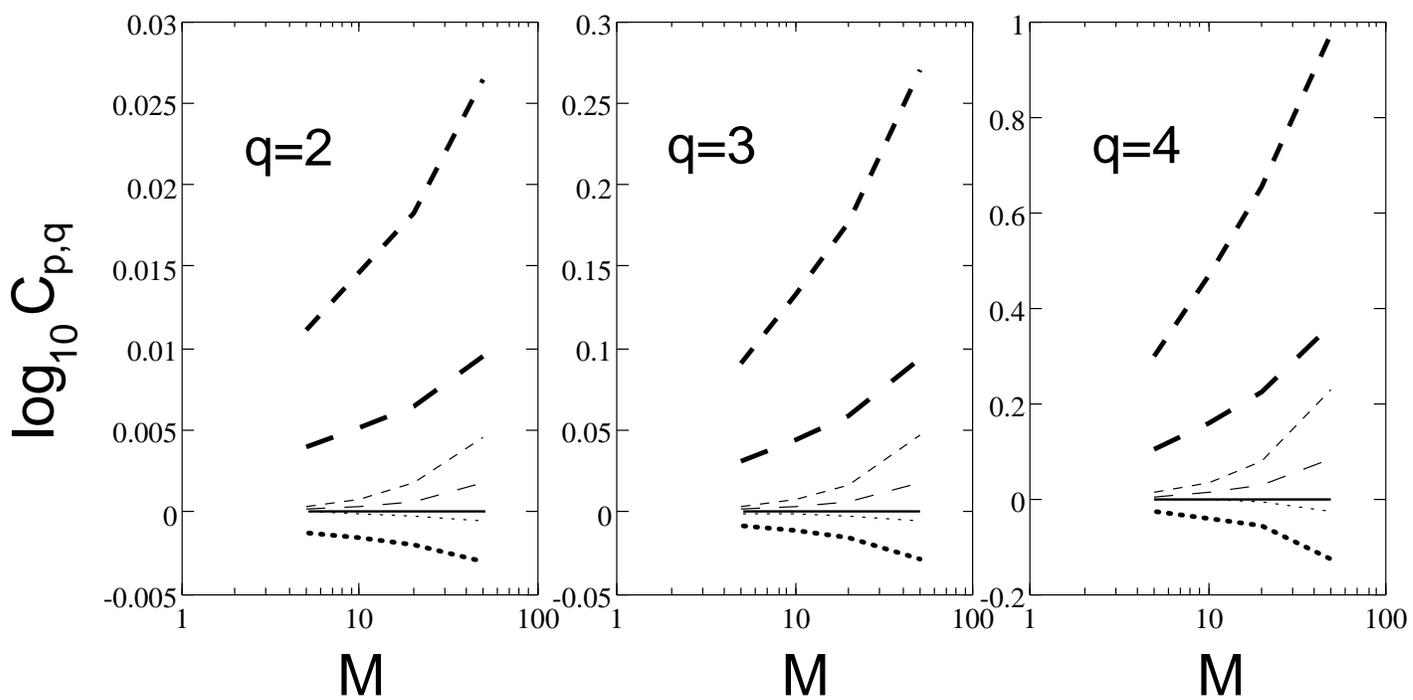
\end{document}